\documentclass[twocolumn,secnumarabic,amssymb, nobibnotes, aps, prx, superscriptaddress]{revtex4-1}
\usepackage{hyperref}
\usepackage{graphics}
\usepackage{graphicx}
\DeclareGraphicsExtensions{.eps}

\begin{document}

\title{Enhanced strain coupling of nitrogen vacancy spins to nanoscale diamond cantilevers}%

\author{Srujan Meesala}%
\thanks{These two authors contributed equally}
\affiliation{John A. Paulson School of Engineering and Applied Sciences, Harvard University, 29 Oxford Street, Cambridge, MA 02138, USA}
\author{Young-Ik Sohn}%
\thanks{These two authors contributed equally}
\affiliation{John A. Paulson School of Engineering and Applied Sciences, Harvard University, 29 Oxford Street, Cambridge, MA 02138, USA}
\author{Haig A. Atikian}%
\affiliation{John A. Paulson School of Engineering and Applied Sciences, Harvard University, 29 Oxford Street, Cambridge, MA 02138, USA}
\author{Samuel Kim}%
\thanks{Current address: Research and Exploratory Development Department, Johns Hopkins University Applied Physics Laboratory, Laurel, MD 20723, USA}
\affiliation{John A. Paulson School of Engineering and Applied Sciences, Harvard University, 29 Oxford Street, Cambridge, MA 02138, USA}
\author{Michael J. Burek}%
\affiliation{John A. Paulson School of Engineering and Applied Sciences, Harvard University, 29 Oxford Street, Cambridge, MA 02138, USA}
\author{Jennifer T. Choy}%
\thanks{Current address:  Draper Laboratory, 555 Technology Square, Cambridge, MA 02139, USA}
\affiliation{John A. Paulson School of Engineering and Applied Sciences, Harvard University, 29 Oxford Street, Cambridge, MA 02138, USA}
\author{Marko Lon\v{c}ar}%
\email{loncar@seas.harvard.edu}
\affiliation{John A. Paulson School of Engineering and Applied Sciences, Harvard University, 29 Oxford Street, Cambridge, MA 02138, USA}
\date{October 18, 2015}%

\begin{abstract}
Nitrogen vacancy (NV) centers can couple to confined phonons in diamond mechanical resonators via the effect of lattice strain on their energy levels. Access to the strong spin-phonon coupling regime with this system requires resonators with nanoscale dimensions in order to overcome the weak strain response of the NV ground state spin sublevels. In this work, we study NVs in diamond cantilevers with lateral dimensions of a few hundred nm. Coupling of the NV ground state spin to the mechanical mode is detected in electron spin resonance (ESR), and its temporal dynamics are measured via spin echo. Our small mechanical mode volume leads to a 10-100$\times$ enhancement in spin-phonon coupling strength over previous NV-strain coupling demonstrations. This is an important step towards strong spin-phonon coupling, which can enable phonon-mediated quantum information processing and quantum metrology.
\end{abstract}
\maketitle

\section{Introduction}

Quantum two level systems (qubits) strongly coupled to mechanical resonators can function as hybrid quantum systems with several potential applications in quantum information science \cite{1402-4896-2009-T137-014001, PhysRevLett.105.220501, PhysRevA.84.042341}. The physics of these systems can be well described with the tools of cavity quantum electrodynamics (cQED), in which an atom is strongly coupled to photons in an electromagnetic cavity. Such qubit-mechanical mode interactions are key ingredients for quantum logic with ion traps \cite{Leibfried}, and have been used to generate non-classical states of a mechanical resonator coupled to a superconducting qubit \cite{Cleland}. Furthermore, phonons or mechanical vibrations couple to a wide variety of well-studied quantum systems, and are therefore, considered a promising means to coherently interface qubits across disparate energy scales \cite{1402-4896-2009-T137-014001, PhysRevLett.105.220501, PhysRevA.84.042341}. In the context of solid state emitters, mechanical hybrid systems were first proposed \cite{PhysRevLett.92.075507}, and subsequently demonstrated \cite{PhysRevLett.105.037401, QDNanowire} with the electronic states of quantum dots. Such a mechanical hybrid quantum system with negatively charged nitrogen vacancy (NV(-), hereafter referred to as NV) centers in diamond, in particular, would benefit from their long spin coherence times \cite{T2.1sec}. It has been proposed that in the strong spin-phonon coupling regime, phonons can be used to mediate quantum state transfer, and generate effective interactions between NV spins \cite{RablNEMS}. Strong coupling of an NV spin ensemble to a mechanical resonator can also be used to generate squeezed spin states \cite{PhysRevLett.110.156402}, which can enable high sensitivity magnetometry \cite{PhysRevLett.109.253605}. 

Seminal experiments on coupling NV spins to mechanical oscillators relied on magnetic field gradients \cite{Kolkowitz30032012, doi:10.1021/nl300775c, NVnanowire}. More recently, owing to the development of single crystal diamond nanofabrication techniques \cite{BirgitFab, doi:10.1021/nl204449n, doi:10.1021/nl302541e, :/content/aip/journal/apl/101/16/10.1063/1.4760274, DegenCantilever, doi:10.1021/acs.nanolett.5b01346, BarclayWaveguide}, the effect of lattice strain on the NV ground state spin sublevels has been exploited to couple NVs to mechanical modes of diamond cantilevers \cite{AniaCantilever, PhysRevLett.113.020503, PatrickStrongDriving}, and bulk acoustic wave resonators fabricated on diamond \cite{PhysRevLett.111.227602, MacQuarrie:15, GregDD}. Strain-mediated coupling is experimentally elegant since its origin is intrinsic to a monolithic device, and it does not involve functionalization of mechanical resonators, or precise and stable positioning of magnetic tips very close to a diamond chip. However, current demonstrations are far from the strong coupling regime due to the small spin-phonon coupling strength provided by strain from relatively large mechanical resonators. In this work, we present an important step towards strong coupling by incorporating photostable NVs in a diamond cantilever with nanoscale transverse dimensions, and demonstrate a single phonon coupling rate of $\sim{2}$ Hz from dispersive interaction of NV spins with the resonator. This is a $\sim 10-100\times$ improvement over existing NV-strain coupling demonstrations. In our experiments, we first detect the effect of driven cantilever motion on NVs as a broadening of their electron spin resonance (ESR) signal, and through follow-up measurements, establish this to be strain-mediated coupling to the mechanical mode of interest. Subsequently, we use spin echo to probe the temporal dynamics of NVs in the cantilever, and precisely measure the spin-phonon coupling rate. In the conclusion, we discuss subsequent device engineering options to further improve this coupling strength by $\sim{100} \times$, and reach the strong coupling regime. 

\section{Requirements for strong spin-phonon coupling}
In analogy with atom-photon interactions in cQED, the key requirement for applications that rely on strong qubit-phonon coupling is that the co-operativity of the interaction exceed unity \cite{Kolkowitz30032012} 

\begin{equation}
C = \frac{g^2}{n_{th}\kappa\gamma} > 1
\label{cooperativity}
\end{equation}

Here, $g$ is the single phonon coupling rate, $n_{th}$ is the thermal phonon occupation of the mechanical mode of interest, $\kappa$ is the intrinsic mechanical damping rate, and $\gamma$ is the qubit dephasing rate. For strain-mediated linear coupling, the single phonon coupling rate is given by $g=d\epsilon_{ZPM}$, where $\epsilon_{ZPM}$ is the strain due to zero point motion, and $d$ is the strain susceptibility, an intrinsic property of the qubit. The spin triplet ground state of the NV has a relatively small $d\approx 10-20$ GHz/strain \cite{AniaCantilever, PhysRevLett.113.020503}, since the three spin sublevels share the same orbital wavefunction. The effect of strain on these levels is proposed to be a filtered down effect from stronger perturbations to the orbitals themselves, particularly from spin-orbit coupling to the excited state \cite{PhysRevB.85.205203}, and a change in spin-spin interaction energy in the deformed ground state orbital \cite{PhysRevLett.112.047601}. Thus, engineering the mechanical mode to provide large $\epsilon_{ZPM}$ is essential to achieve large $g$. For instance, for the fundamental out-of-plane flexural mode of a cantilever of width $w$, thickness $t$, and length $l$, we can use Euler-Bernoulli beam theory \cite{EBtheory} to show that 

\begin{equation}
\epsilon_{ZPM} \propto \frac{1}{\sqrt{l^3 w}}
\label{eps_zpm}
\end{equation}

This sharp inverse scaling of $\epsilon_{ZPM}$ with cantilever dimensions highlights the importance of working with small resonators. It is analogous to the $1/\sqrt{V_{eff}}$ scaling of the single photon Rabi frequency in cQED, where $V_{eff}$ is the electromagnetic mode volume. To achieve the strong coupling condition in Eq. \ref{cooperativity}, assuming an NV spin coherence time $T_2 = 100$ ms, a mechanical quality factor $Q=\omega/\kappa=10^6$, and cryogenic operation temperatures (4K or lower), cantilevers of width $w \sim 50-100$ nm and length $l \sim 1$ $\mathrm{\mu}$m (corresponding mechanical frequency, $\omega_m \approx$ few hundred MHz) that provide $g \approx$ few hundred Hz are required. Towards this end, our devices in this work have $w$ of the order of a few hundred nm. This is typically the length scale at which proximity to surfaces begins to deteriorate the photostability of the NV charge state.

\section{Device fabrication and experimental setup}

\begin{figure}
\includegraphics[width=\columnwidth]{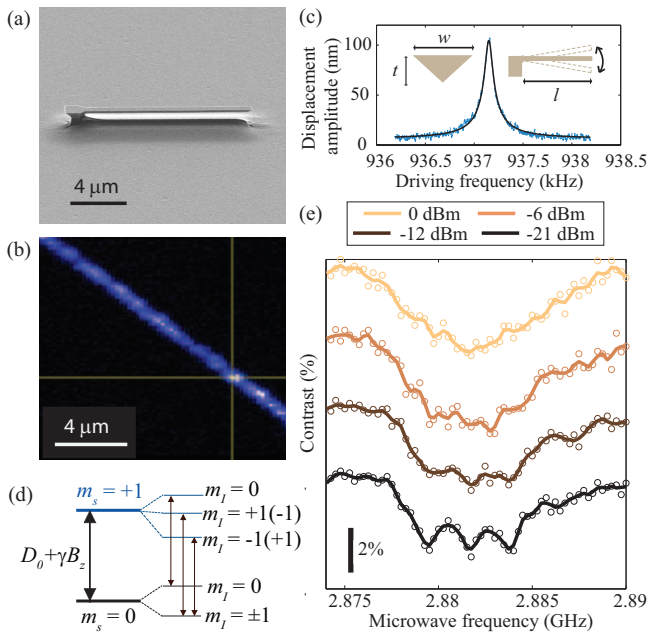}
\caption{(a) Representative scanning electron microscope (SEM) image of the angle-etched diamond cantilevers used. (b) Representative confocal microscope scan of a section of the cantilever showing fluoresecence from NV centers. (c) Driven response of the fundamental out-of-plane flexural mode (right inset) of the triangular cross section (left  inset) cantilevers studied in this work. For this  particular device, we have $w$=580 nm, $t$=170 nm, and $l$=19 $\mathrm{\mu}$m. The mode frequency is 937.2 kHz, and it has a Q-factor of ~10,000. Measurements were taken in high vacuum (1e-5 torr) at room temperature. (d) Hyperfine structure of the $m_s = 0$ to $m_s = +1$ electron spin transition in the NV ground state indicating the three allowed microwave transitions. (e) AC strain induced broadening of the $m_s = 0$ to $m_s = +1$ hyperfine transitions near the clamp of the cantilever with gradually increasing mechanical amplitude. The mechanical mode is inertially driven at its resonance frequency with a piezo stack in all measurements. Open circles indicate measured data, and smoothed solid lines serve as a guide to the eye. Legend shows values of piezo drive power for each measurement. 0 dBm of drive power corresponds to an amplitude of $559\pm 2$ nm at the tip of the cantilever.}
\label{fig1}
\end{figure}

Incorporating photostable NV centers close to surfaces \cite{:/content/aip/journal/apl/96/12/10.1063/1.3364135, doi:10.1021/acs.nanolett.5b00457}, particularly in nanostructures with small transverse dimensions such as nanophotonic cavities \cite{doi:10.1021/nl402174g, PhysRevLett.109.033604}, has been found to be a considerably challenging task in recent years. To prevent charge state blinking and photo-ionization of NVs under optical excitation \cite{PhysRevLett.110.167402}, high quality surfaces with low defect density, and appropriate surface termination are necessary. Recent advances in annealing and surface passivation procedures \cite{doi:10.1021/nl404836p} have significantly improved the ability to retain photostable NV centers generated by ion implantation even after fabrication of nanostructures around them \cite{DeLeonNano}. Using these techniques in combination with our angled reactive ion etching (RIE) fabrication scheme \cite{doi:10.1021/nl302541e} (details discussed in Appendix A), we were able to generate photostable NVs in diamond nano-cantilevers with a triangular cross section (Fig. \ref{fig1}(a,b)). Previously, we have demonstrated high Q-factor mechanical modes (Q approaching 100,000) with frequencies ranging from $<$1 MHz to tens of MHz in cantilevers, and doubly clamped nanobeams fabricated using the same angled etching scheme \cite{:/content/aip/journal/apl/103/13/10.1063/1.4821917}. 

Our measurements are carried out at high vacuum ($10^{-5}$ torr), and room temperature in a vacuum chamber with a view port underneath a homebuilt scanning confocal microscope for addressing NV centers. Microwaves for ESR measurements are delivered with a wire bond positioned close to the devices of interest. The diamond chip is mounted on a piezo actuator for resonant actuation of cantilevers. Mechanical mode spectroscopy performed via optical interferometry \cite{:/content/aip/journal/apl/86/1/10.1063/1.1843289} is used to characterize the modes of the cantilevers. For the experiments described in this paper, we used a triangular cross-section cantilever with $w$=580 nm, $t$=170 nm, and $l$=19 $\mathrm{\mu}$m. The mechanical mode of interest (Fig. \ref{fig1}(c)) is the out-of-plane flexural mode, which was found to have a frequency, $\omega_m = 2\pi \times 937.2$ kHz, and a quality factor, $Q \sim 10,000$. 

\section{AC strain induced ESR broadening}

The effect of lattice strain on the $S=1$ ground state of the NV center has been described by the Hamiltonian \cite{AniaCantilever, PhysRevLett.113.020503, PhysRevB.85.205203, PhysRevLett.110.156402} -

\begin{equation}
H = D_0 S_z^2 + \gamma \mathbf{S\cdot B} + d_{\parallel}\epsilon_{zz}S_z^2 - \frac{d_{\perp}}{2} \left[\epsilon_{+} S_{+}^2 + \epsilon_{-} S_{-}^2 \right] 
\label{Hstrain}
\end{equation}

Here $S_i$ are the $S=1$ Pauli spin operators. $D_0=2.87$ GHz is the zero-field splitting between $m_s=0$ and $m_s=\pm 1$ levels due to spin-spin interaction, $\gamma=2.8$ MHz/G is the gyromagnetic ratio for the NV ground state. At small B-fields ($\ll D_0/\gamma$), the NV-axis is the spin quantization axis ($z-$axis in the above Hamiltonian).  $d_{\parallel}$ and $d_{\perp}$ are respectively the axial and transverse strain susceptibilities defined with reference to the NV-axis. 
$\epsilon_{ii}$ are the diagonal strain tensor components defined in the basis of the NV, and $\epsilon_{\pm} = \epsilon_{xx} \pm i\epsilon_{yy}$. This Hamiltonian neglects the shear components of the strain tensor, which are relatively insignificant for the chosen mechanical mode. The perturbative strain terms lead to frequency shifts in the $m_s=\pm 1$ levels respectively given by

\begin{equation}
\Delta\omega_{\pm} = d_{\parallel}\epsilon_{\parallel} \pm \sqrt{\left( \gamma B_z \right)^2 + \left( d_{\perp} \epsilon_{\perp} \right)^2}
\label{freq_shifts}
\end{equation}

Here, $\epsilon_{\perp}$ denotes the total transverse strain $\sqrt{\epsilon_{xx}^2+\epsilon_{yy}^2}$. Physically, Eq. \ref{freq_shifts} reveals that axial strain leads to a linear modification of the zero-field splitting, while transverse strain mixes the $m_s=\pm 1$ states, thereby causing a quadratic splitting between them. In a mechanical resonator driven at the frequency $\omega_m$, the local strain components $\epsilon_{\parallel}$ and $\epsilon_{\perp}$ oscillate at the frequency $\omega_m$. The classical effect of driving the mechanical mode on the NV ground state is frequency modulation of the two transitions between $m_s=0$ and $m_s=\pm 1$ levels. From the strain susceptibilities measured in [8], and finite element calculations on our structures, we anticipate a frequency modulation comparable to the ESR linewidth, when the mechanical mode is driven to an amplitude of $\approx$500 nm.

At the chosen nitrogen ion implantation density, we expect $\sim{10}$ NV centers within our confocal laser spot. ESR measurements are performed on such an NV ensemble at a fixed position in the cantilever, and simultaneously, the flexural mode shown in Fig. 1(c) is driven by supplying an RF voltage to the piezo actuator at the resonance frequency $\omega_m$. A small static magnetic field $B_z$= 4 G is applied with a bar magnet placed outside the cryostat, and only the $m_s=0$ to $m_s=+1$ transition is probed. The external magnetic field is aligned exactly vertically to ensure that all four NV classes experience the same projection $B_z$ along their respective axes. The cantilever itself is fabricated such that its long axis is aligned to the $<100>$ crystal axis to within a few degrees as determined by electron back scatter diffraction (EBSD). As a result, all four NV classes are symmetrically aligned with respect to the dominant strain component of the flexural mode, which occurs along the cantilever long axis. Thus, at a given location in the cantilever, all four NV classes experience the same axial and transverse strain amplitudes, and hence experience identical transition frequency modulation. Effects of inhomogeneous coupling strength due to implantation straggle, and varying lateral position within the confocal laser spot are addressed in Appendix B. Low microwave power was used to prevent power broadening, and retain near native linewidths in the ESR. 

Fig. \ref{fig1}(e) shows ESR spectra at the same location in the cantilever for progressively increasing mechanical amplitude. At the lowest piezo drive power of -21 dBm, we observe three dips spaced equally by 2.2 MHz corresponding to the hyperfine structure arising from interaction between the NV electron spin and the $^{14}$N nuclear spin (Fig. \ref{fig1}(d)). This was found to be identical to the ESR spectrum with no piezo drive (not shown). For each of the hyperfine transitions, we measure a linewidth of $\approx$2 MHz. As the piezo drive power is increased to -12 dBm, we observe a broadening of the hyperfine features to the point where the hyperfine structure is barely resolvable. At -6 dBm, the hyperfine structure is washed out, and at 0 dBm, the overall ESR dip is even broader. Such broadening of the ESR signal with progressively larger mechanical amplitude is expected, since the measurement sequence involves dwelling at each microwave frequency sample for many ($> 10^6$) cycles of the mechanical oscillation period. As a result, we would expect to average over the AC modulation of the microwave transition, and detect an overall broadening determined by the modulation amplitude.  

\begin{figure}
\includegraphics[width=\columnwidth]{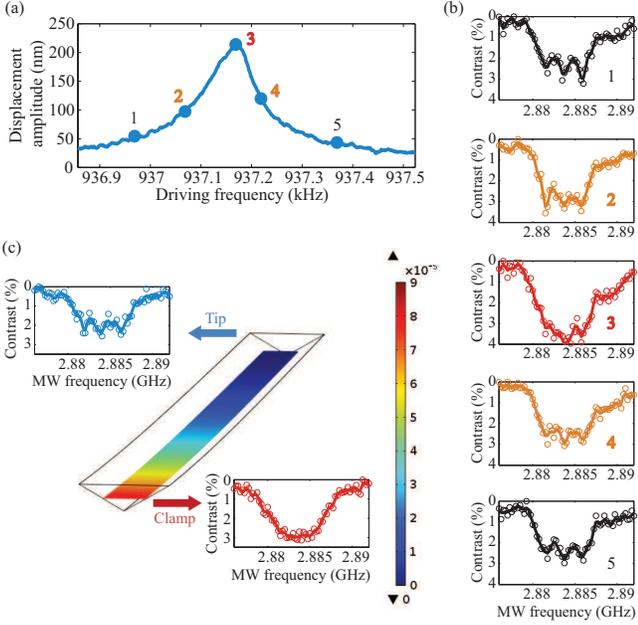}
\caption{(a) Driven response of the cantilever at a piezo drive power of -12 dBm. The drive frequencies used for frequency dependent broadening measurements are indicated with numbers 1-5. (b) ESR spectra at the same location in the cantilever at mechanical drive frequencies 1-5 indicated in (a). (c) ESR spectra at the tip and clamp of the cantilever for -6 dBm drive power. Strain profile of the mechanical mode from an FEM simulation for the corresponding displacement amplitude is also shown. Open circles in each ESR spectrum are measured data, and smoothed lines serve as a guide to the eye}
\label{fig2}
\end{figure}

To verify that the ESR broadening arises from the mechanical mode, we perform the same measurement at a fixed drive power of -12 dBm, and multiple drive frequencies around resonance. The slight asymmetry in the driven mechanical response (Fig. \ref{fig2}(a)) can be attributed to the onset of a Duffing-type nonlinearity at this drive power. When driven far off the mechanical resonance as in points 1 and 5 in Fig. \ref{fig2}(b), the ESR spectrum retains three clear hyperfine dips with linewidths close to the native linewidth. At smaller detunings as in points 2 and 4 in \ref{fig2}(b), we observe a broadening of the individual hyperfine features. When driven exactly on resonance as in point 3 in Fig. \ref{fig2}(d), the hyperfine features are on the verge of being washed out. Thus, the ESR broadening effect follows the frequency response of the mechanical mode. To further confirm that this is a strain-induced effect, we repeat the measurement at multiple points along the length of the cantilever for a fixed piezo drive power of -6 dBm. From the strain profile of the flexural mode (Fig. \ref{fig2}(c)), we expect a roughly linear variation in AC strain amplitude from its maximum value near the clamp of the cantilever to zero at the tip of the cantilever. This effect is observed in the form of ESR broadening for NVs near the clamp, and retention of native linewidths for NVs at the tip (Fig. \ref{fig2}(c)).   

\section{Temporal dynamics of the mechanically driven spin}
The ESR broadening measurements in Figs. \ref{fig1}(e) and Fig. \ref{fig2} provide strong evidence of strain from the driven mechanical mode coupling to the NV spin. From the washing out of hyperfine structure in the measurements, we can deduce driven coupling rates of the order of the hyperfine splitting (2.2 MHz). In order to probe the temporal dynamics of the NV spin due to mechanical motion, and precisely measure the coupling strength, we employ spin echo measurements. It has been shown in previous demonstrations that the two distinct modes of level shifts generated by axial and transverse strain can be used to achieve dispersive \cite{AniaCantilever} and resonant interactions \cite{PhysRevLett.111.227602, GregDD, MacQuarrie:15, PatrickStrongDriving} of the spin with mechanical motion, respectively. In our work, the  frequency of our mechanical mode ($\approx$1 MHz) is smaller than the ESR linewidth, and we will focus on the dispersive regime provided by axial strain. We apply a moderate static magnetic field, and suppress the effect of transverse strain to first order as evinced by Eq. \ref{freq_shifts}. In this regime, if we work with the effective qubit defined by the $m_s=0$ and $m_s=+1$ levels, driven motion of the mechanical resonator can modulate the phase of our effective qubit at the frequency $\omega_m$, analogous to an AC magnetic field. This is described by the time dependent Hamiltonian  

\begin{equation}
H_{int}(t) = 2\pi G \mathrm{cos}\left(\omega_m t + \phi \right) \sigma_z
\end{equation}

Here $G=d_{\parallel}\epsilon_{zz}$ is the AC strain coupling rate from the driven motion, $\phi$ is an arbitrary phase offset, and $\sigma_z$ is the corresponding $S=1/2$ Pauli spin operator.

\begin{figure}
\includegraphics[width=\columnwidth]{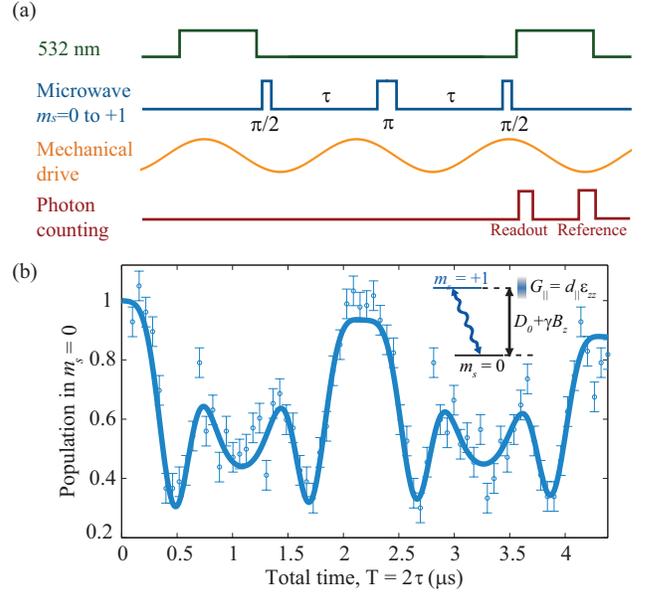}
\caption{(a) Experimental pulse sequence for spin echo measurement of dispersive spin-cantilever interaction due to axial strain
(b) Spin echo signal from NVs in the cantilever at a piezo drive power of 0 dBm (tip amplitude of 559$\pm$2 nm) for the mode at $\omega_m$ = 923.4 kHz, showing two periods of the modulation due to axial strain coupling. The solid line is a fit to Eq. \ref{bessel}. Vertical error bars correspond to photon shot noise in the measurement. Inset shows schematic of dispersive interaction between the qubit and mechanical mode due to axial strain.}
\label{fig3}
\end{figure}

In our spin-echo measurements, we apply an external static magnetic field $B_z$= 27 G. As in the case of ESR measurements, the magnetic field is aligned to ensure equal Zeeman splittings for all four NV orientations. Our experimental sequence is shown in Fig. \ref{fig3}(a), wherein the piezo drive signal, and hence the strain field, has an arbitrary phase $\phi$ with respect to the microwave pulses that varies over multiple iterations of the sequence. The spin echo signal obtained from this measurement (Fig. \ref{fig3}(b)) at a piezo drive power of 0 dBm shows a periodicity corresponding to twice the time period of the mechanical mode. The theoretically expected spin echo signal in this measurement has the form of a zero order Bessel function with a periodic argument \cite{Kolkowitz30032012, AniaCantilever}.

\begin{equation}
p(2\tau) = \frac{1}{2} \left[ 1 + e^{-\left( 2\tau/T_2 \right)^3} J_0 \left( \frac{8\pi G}{\omega_m} \mathrm{sin}^2\left( \frac{\omega_m \tau}{4} \right) \right)  \right]
\label{bessel}
\end{equation}

The exponential damping term multiplying the periodic function corresponds to dephasing of the NV electron spin due to interactions with the surrounding $^{13}$C nuclear spin bath in diamond, and $T_2$ is the dephasing time \cite{Childress13102006}. In our experiments, $T_2 \gg$ the mechanical oscillation period ($2\pi/\omega_m$), and the effect of spin decoherence is relatively small. A fit to expression \ref{bessel} yields $\omega_m= 2\pi\times 918.7 \pm 5.6$ kHz, which is in reasonable agreement with the driving frequency of 923.4 kHz used in the experiment. The extracted driven coupling rate $G = 2.10 \pm 0.07$ MHz is of the order of the hyperfine splitting as observed in our ESR broadening measurements.

\begin{figure}
\includegraphics[width=\columnwidth]{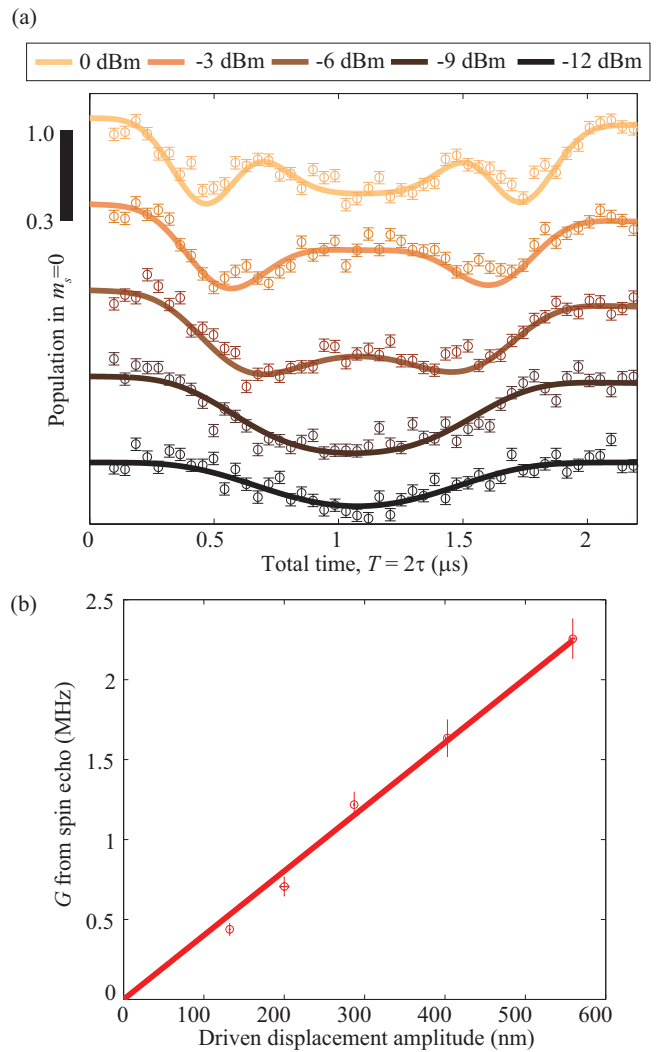}
\caption{(a) Spin echo at the same location in the cantilever for varying piezo drive powers, offset along y-axis. One period of the signal is plotted in each case. Solid lines are fits to the zero order Bessel function form indicated in the text, and vertical error bars correspond to photon shot noise. The scale bar on the side is a guide for the $y$-axis indicating maximum (1.0) and minimum (0.3) possible population in $m_s=0$ as dictated by Eq. \ref{bessel}. The legend on right side indicates the piezo drive powers used for each measurement. (b) Variation of driven spin-phonon coupling rate due to axial strain ($G$) with the calibrated displacement amplitude of the mechanical mode. The five data points correspond to the piezo drive powers used in plot (a) in increasing order. Vertical error bars correspond to the error in each $G$ estimate from fitting to the spin echo function given by Eq. \ref{bessel}. Solid line is a linear fit, which yields $dG/dx = 4.02 \pm 0.40$ kHz/nm.}
\label{fig4}
\end{figure}

Finally, we performed spin echo measurements at the same location on the cantilever for varying piezo drive powers (Fig. \ref{fig4}(a)). The no-drive spin echo signal (not shown) is flat, indicating that this measurement is not sensitive to the thermal motion of the cantilever mode (estimated to have an amplitude of 0.53 nm). As the cantilever is driven, the spin echo signal begins to show a dip when the evolution time $2\tau$ equals the mechanical oscillation period ($2\pi/\omega_m$). At larger amplitudes, the spin precesses by more than one full rotation on the equator of the Bloch sphere, and we observe higher order fringes within one period of the signal. These drive-power dependent measurements further allow us to verify that the axial strain coupling is linear in the displacement amplitude (Fig. \ref{fig4}(b)). From the linear fit, we infer a displacement sensitivity $dG/dx = 4.02 \pm 0.40$ kHz/nm. By estimating the zero point motion of the mode from its effective mass, this displacement sensitivity yields a single phonon coupling strength, $g = 1.84 \pm 0.18$ Hz for an NV at the clamp of this cantilever. Compared with recent demonstrations of NV-strain coupling, this is about two orders of magnitude larger than that measured in \cite{AniaCantilever}, and an order of magnitude larger than that in \cite{PatrickStrongDriving, PhysRevLett.113.020503}. 

\section{Conclusion and outlook for strong coupling}

In conclusion, we demonstrate nanoscale diamond cantilevers for strain-mediated coupling of NV spins to mechanical resonators. The relatively small dimensions of our devices offer a significant improvement in the single phonon coupling strength compared to previous work with mechanical modes at similar frequencies. Shorter cantilevers will boost $g$ even further according to the scaling in Eq. \ref{eps_zpm}. This will also increase the mechanical frequencies, and allow operation in the sideband resolved regime with access to the resonant spin-phonon interaction provided by transverse strain coupling \cite{PhysRevLett.111.227602, GregDD, MacQuarrie:15, PatrickStrongDriving}. As suggested by our estimates in the introduction, nanostructures that allow strong coupling have mechanical frequencies in the few hundreds of MHz range. Our recent efforts have addressed the goal of developing nanomechanical resonators in single crystal diamond at such high frequencies. Dynamic actuation and transduction of single crystal diamond resonators up to frequencies of 50 MHz was achieved with dielectric gradient forces in \cite{SohnDielectro}, and up to 6 GHz was achieved by using cavity optomechanics in \cite{BurekOMC}. Given these developments, we anticipate that the major engineering challenge for strain-mediated strong spin-phonon coupling will be the ability to maintain photostable NVs in nanostructures with extremely small widths (the current state-of-the-art being nanobeams with $w \approx 200$ nm in \cite{DeLeonNano}). Another challenge is the demand for high Q-factors in the $10^5-10^6$ range from small resonators, which are usually challenging to engineer for high Q \cite{Imboden201489}.

However, these system engineering requirements can be less demanding for magnetometry applications that can benefit from an NV ensemble coupled to a mechanical resonator \cite{PhysRevLett.110.156402}. In particular, collective enhancement from a dense spin ensemble can boost the co-operativity by a factor of the number of spins $N$ \cite{PhysRevLett.102.083602}, allowing one to work with device dimensions more favorable for NV photostability and high mechanical Q-factors. Alternatively, since the framework of strain coupling outlined above is fairly general, the same devices may be used, but with a different qubit, whose energy levels have a larger strain response. Potential candidates include the NV center excited electronic state \cite{PhysRevLett.102.195506, DaviesAndHamer}, and the orbital ground states of the silicon vacancy (SiV) center \cite{PhysRevB.50.14554}, both of which have 4-5 orders of magnitude larger strain susceptibility than the NV ground state spin sublevels. 

\appendix
\section{Sample preparation} 
Single crystal electronic grade bulk diamond chips (4mm x 4mm) from Element Six Ltd are implanted with $^{14}$N ions at an implantation energy of 75 keV, and a dose of $6\times 10^{11}$ /cm$^{2}$. This yields an expected depth of $94 \pm 19$ nm calculated using software from Stopping and Range of Ions in Matter (SRIM). Subsequently, NVs are created by annealing the samples in high vacuum ($< 5\times 10^{-7}$ torr).  The temperature ramp sequence described in \cite{doi:10.1021/nl404836p} is followed with a final temperature of $1200^{o}$C, which is maintained for 2 hours. After the anneal, the samples are cleaned in a 1:1:1 boiling mixture of sulfuric, nitric and perchloric acids to remove a few nm of graphite generated on the surface from the anneal. Cantilevers are then patterned using e-beam lithography, and etched using our angled etching scheme \cite{doi:10.1021/nl302541e}. Post-fabrication, we repeat the tri-acid cleaning treatment to partially repair etch-induced damage, and perform a piranha clean to ensure a predominantly oxygen terminated diamond surface (diagnosed by X-ray photoelectron spectroscopy (XPS)), which is beneficial for NV photostability \cite{doi:10.1021/nl404836p, DeLeonNano}.\\

\section{Ensemble effects}
We address the effect of inhomogeneous coupling strengths in AC strain coupling measurements on an NV ensemble. The width of our confocal laser spot $\approx 560$ nm is about forty times smaller than the length of the cantilever (19 $\mathrm{\mu}$m). Taking into account the roughly linear variation of strain along the length of the cantilever, we expect a $\approx 2\%$ variation in coupling strength within the confocal spot due to lateral distribution of NVs. This is less than the order of the error in the fitted estimate for $G$. Now, we consider the more significant effect of ion-implantation straggle, which is expected to be $\approx 20 \%$ for our chosen NV depth from SRIM. Upon fitting the experimental spin echo signal to the formula in Eq. \ref{bessel} convolved with a 20$\%$ Gaussian straggle in $G$, we noticed that our estimate for $G$ did not change to within the error bars. We believe that this is because the level of photon shot noise in our measurement ($\pm 0.05$ error in spin population estimates) does not allow us to ultimately resolve the effect of any inhomogeneity in $G$ across the ensemble.

\begin{acknowledgments}
The authors thank Ruffin Evans, Nathalie de Leon, Kristiaan de Greve, and Yiwen Chu for sharing diamond annealing and surface treatment procedures, and acknowledge Arthur Safira, Vivek Venkataraman, and Mikhail Lukin for helpful discussions. This work was supported by DARPA QuASAR (Award No. HR0011-11-C-0073), STC Center for Integrated Quantum Materials (NSF Grant No. DMR-1231319), and ONR MURI on Quantum Optomechanics (Award No. N00014-15-1-2761). H.A. Atikian and M.J. Burek were supported in part by the Natural Science and Engineering Council (NSERC) of Canada, and the Harvard Quanum Optics Center (HQOC). This work was performed in part at the Center for Nanoscale Systems (CNS), a member of the National Nanotechnology Infrastructure Network (NNIN), which is supported by the National Science Foundation under NSF award no. ECS-0335765. CNS is part of Harvard University.
\end{acknowledgments}

\end{document}